\documentstyle[12pt]{article}
\font\titlefont=cmbx10 scaled \magstep3

\def\lprox{\mathrel{\raise .3ex\hbox{$<$\kern-
.75em\lower1ex\hbox{$\sim$}}}}

\def\gprox{\mathrel{\raise .3ex\hbox{$>$\kern-
.75em\lower1ex\hbox{$\sim$}}}}

\begin{document}

\begin{center}
{\titlefont Some Thoughts on Energy Conditions and 
Wormholes} \\
\vskip .7in
Thomas A. Roman \footnote
{email: roman@ccsu.edu} \\
\vskip .2in
Department of Mathematical Sciences\\
Central Connecticut State University\\
New Britain, CT 06050  USA\\
\end{center}

\vspace*{1cm}
\begin{abstract}
This essay reviews some of the recent 
progress in the area of energy conditions and wormholes. Most of the discussion centers 
on the subject of ``quantum inequality'' restrictions on negative energy. 
These are bounds on the magnitude and duration of negative energy which 
put rather severe constraints on its possible macroscopic effects. Such 
effects might include the construction of wormholes and warp drives for 
faster-than-light travel, and violations of the second law of thermodynamics. 
Open problems and future directions are also discussed.
\end{abstract}
\newpage

\baselineskip=24pt

\section{Introduction}

This paper focuses primarily on ``quantum inequality'' restrictions 
on negative energy. Although some aspects of wormholes are briefly discussed, 
for a recent summary of work on the subject since 1995, the reader is referred 
to the Introduction of the paper by Lemos, Lobo, and Quintet de Oliveira \cite{LLQ}. 
For a good recent review of the current status of the 
related issue of time machines, see Visser's article in the 
Hawking 60th Birthday volume \cite{V-HBV}. A more technical review of 
quantum inequalities, which in many respects is complementary to the present one, 
is by Fewster \cite{Few-Rew}. The main purpose of the current review is to give 
the reader the physical flavor of the results discussed.

\subsection{Energy Conditions}
\label{subsec:EC}

In solving the Einstein equations, one usually proceeds in the 
following way. Assume a ``physically reasonable'' distribution 
of mass-energy, e.g., a cloud of gas particles, EM field, etc. 
Solve (i.e., integrate) the Einstein field equations to find the 
spacetime geometry produced by this distribution. This is typically 
a difficult problem unless there is a high degree of symmetry 
(e.g., spherical or axial symmetry). A more fundamental problem is: 
how does one define ``physically reasonable''?

Historically, this notion has been defined by imposing ``energy conditions'' on the 
stress-energy tensor. The weakest such local condition assumed is the 
``weak energy condition (WEC)'', 
\begin{equation} 
T_{\mu\nu}\, u^\mu\, u^\nu \, \geq 0 \,, 
\label{eq:wec} 
\end{equation}
for all timelike observers, 
where $u^\mu$ is the observer's four-velocity. By continuity, 
the condition also holds for null vectors (the ``null energy condition (NEC)''):
\begin{equation} 
T_{\mu\nu}\, K^\mu\, K^\nu \, \geq 0 \,, 
\label{eq:nec} 
\end{equation}
where $K^\mu$ is any null vector.  
These conditions essentially say that the 
energy density is greater than or equal to zero, for all observers. 
More recently, weaker ``averaged'' versions of these conditions have been proposed, 
notably, the ``averaged weak energy condition (AWEC)''
\begin{equation} 
\int_{-\infty}^{\infty} T_{\mu\nu}\, u^\mu\, u^\nu \, d\tau \geq 0 \,, 
\label{eq:awec} 
\end{equation} 
where the average is taken over a timelike {\it geodesic}, $u^\mu$ 
is the tangent to the geodesic, and $\tau$ is the proper time; 
and the ``averaged null energy condition (ANEC)\footnote{Unlike in the case of the 
local conditions, the ANEC cannot in general be obtained from the AWEC. 
They are separate assumptions.}'' 
\begin{equation} 
\int_{-\infty}^{\infty} T_{\mu\nu}\, K^\mu\, K^\nu \, d\lambda \geq 0 \,, 
\label{eq:anec} 
\end{equation}
where the average is taken over a null geodesic, $K^\mu$ 
is the affinely parameterized tangent to the geodesic, and 
$\lambda$ is an affine parameter \cite{T,C-E,Galloway,TR-86,B87,TR-88,K,WY91,Y95,FW,Verch_ANEC}. 
(This is by no means an exhaustive list of 
energy conditions, but it is adequate for our discussion.)
These conditions are obeyed by all {\it observed} forms of classical matter 
and energy. Energy conditions tell us what are ``physically reasonable'' 
distributions of mass-energy, which in turn tell us what are 
physically reasonable spacetime geometries. They are also crucial in 
proving many theorems using global techniques in general relativity, 
such as singularity theorems and the topological censorship theorem \cite{FSW}. 
However, the energy conditions are not derivable from GR. 

Perhaps Einstein had this weakness in mind when, 
with regard to the structure of his field equations, he said: 
``But it [general relativity] is similar to a building, 
one wing of which is made of fine marble (left part of the equation), but 
the other wing of which is built of low grade wood (right side of the equation). 
The phenomenological representation of matter is, in fact, only a crude 
substitute for a representation which would do justice to all known 
properties of matter \cite{AE}.''

\subsection{Quantum Violations}
\label{sec:QV}
It has been known for some time that the WEC and NEC 
are violated even for free quantum fields in Minkowski spacetime \cite{EGJ} 
(e.g. squeezed vacuum states \cite{SY}). The AWEC and ANEC are satisfied 
for free fields in Minkowski spacetime \cite{K,FW} but are 
violated in some curved spacetimes (e.g., Boulware vacuum state \cite{V-BV}). The 
AWEC can be violated in flat spacetimes with boundaries (e.g., for a static observer 
between a pair of Casimir plates \cite{C}), or compact spatial 
dimensions \cite{K}. 

One might ask what are the possible reasons for the existence of 
negative energy in nature? Two thoughts come to mind. The first is the 
stability of flat spacetime. We know from quantum field theory (QFT) that 
the energy density undergoes quantum fluctuations even in the vacuum. 
In order for the expectation value of the stress-tensor to be zero in the 
Minkowski vacuum state, the sign of the energy density must be allowed to 
fluctuate both negatively and positively \cite{LF}. The second is the discovery by Hawking 
that black holes evaporate \cite{H75}. This process can be viewed as a consequence of a flow 
of negative $(-)$ energy down the hole, which pays for the positive $(+)$ energy Hawking flux 
seen at infinity. The emission of radiation from a black hole is what allows it 
to have a meaningfully defined temperature, which in turn allows it (in principle) 
to come into thermal equilibrium with its surroundings. From this follows the beautiful 
and compelling unification of the laws of black hole mechanics with the laws of 
thermodynamics. Therefore, it could be argued that the existence of $(-)$ energy 
is crucial for the consistency of the behavior of black holes with the laws of 
thermodynamics.

Although QFT allows its existence, if there are 
no constraints on the manipulation of $(-)$ energy the 
consequences could be a possible Pandora's box of dramatic 
macroscopic effects. Such effects might include: violation of the 
second law of thermodynamics, violation of ``cosmic censorship'' and the destruction of 
black holes (i. e., the production of naked singularities), singularity avoidance 
inside black holes, breakdown of ``positive mass'' theorems in GR, traversable wormholes, 
warp drive, and time travel. 

If there are no restrictions on $(-)$ energy, then the 
semiclassical Einstein equations would have almost no physical content. 
One could produce ``designer spacetimes.'' Write down any spacetime geometry you 
like, ``plug it in'' to the Einstein equations, and find the corresponding 
mass-energy distribution which generates that geometry. Since {\it any} solution 
of the Einstein equations corresponds to {\it some} distribution of mass-energy, 
with no restrictions (such as some form of energy condition) you can get anything you like. 
Therefore a key fundamental question is: does QFT impose any constraints on $(-)$ energy? 

\subsection{Quantum Inequalities}
\label{sec:QIs} 
Fortunately (or unfortunately, depending on one's point of view) 
the answer to the above question is yes. Over the last decade much work 
has been done in proving so-called ``quantum inequalities (QIs)'', originally introduced by 
Ford \cite{F78}. These 
are restrictions on the magnitude and duration of $(-)$ energy densities and fluxes. 
Let $\rho = \langle\Psi|\rho|\Psi\rangle$ be the (renormalized) expectation value of the energy 
density of a free quantized field in the rest frame of an arbitrary inertial observer, 
in an arbitrary (Hadamard) quantum state $|\Psi\rangle$. Let $t$ be 
the proper time measured along the observer's worldline, and  
$f(t)$ be a ``sampling function'', i.e., a smooth peaked function 
with characteristic width $\tau_0$, and whose integral is 1. 
The QIs have the form
\begin{equation}
\hat \rho = \int_{-\infty}^{\infty} \rho \, f(\tau) \, d\tau
\geq - \,\frac{C \, \hbar}{c^3 \, \tau_0^4} \,,
\label{eq:QI}
\end{equation}
for {\it all} choices of ``sampling time'' $\tau_0$, and where $C \ll 1$, is a 
constant which depends on the choice of sampling function. Here $\hbar$ is Planck's 
constant and $c$ is the speed of light.

Physically, the QIs say that the more negative the energy density is in some time 
interval, the shorter the duration of the interval. 
An inertial observer cannot see arbitrarily large $(-)$ 
energy densities which last for arbitrarily long periods of time. 
QI bounds have now been derived for free fields, including 
the electromagnetic field, the Dirac field, the 
massless and massive scalar fields, massive spin-one, and Rarita-Schwinger 
fields 
using {\it arbitrary} (smooth) sampling functions 
\cite{F78,F91,FR95,FR97,PF971,PFGQI,FE,Vollick1,Vollick2,FT,FLAN1,FLAN2,Fewster,FV,FP,FM,YW}. 
These constraints have the form of an uncertainty principle-type bound. However, 
the uncertainty principle was {\it not} used as input in their derivations. These are 
rigorous mathematical bounds which are directly derivable from QFT. In some sense, they are 
more powerful than the AWEC. For example, the latter implies that an inertial observer who 
initially encounters some $(-)$ energy must encounter compensating $(+)$ energy {\it at 
some time} in the future. By contrast, the QIs say that the inertial observer 
must encounter the compensating $(+)$ energy {\it no later than} a time $T$, which 
is inversely proportional to the magnitude of the initial $(-)$ energy. Furthermore, 
in Minkowski spacetime, the AWEC can be derived as simply the infinite sampling time limit 
of the QIs.

The QIs are extraordinarily rich. For example, they lead to an additional phenomenon 
known as ``quantum interest.'' This effect implies that an initial burst of $(-)$ energy 
must be followed by a subsequent burst of $(+)$ energy within a maximally 
allowed time interval, which is inversely proportional to the magnitude of the 
initial $(-)$ energy burst. In addition, the magnitude of the $(+)$ energy 
burst must {\it exceed} that of the $(-)$ energy burst by an amount which grows 
as the time separation between the bursts increases. One can think of $(-)$ 
energy as an energy ``loan'' which must be paid back (by $(+)$ energy ) 
with an ``interest'' which grows with the magnitude and duration of the ``debt'' 
\cite{FR99,Pre,FT00,TW}. 

The quantum inequalities do not limit the existence of $(-)$ energy, per se, but the 
arbitrary {\it separation} of $(-)$ energy from its $(+)$ energy counterpart. This in turn 
foils various schemes to isolate $(-)$ energy from $(+)$ energy. For example, 
consider a burst of $(-)$ energy followed by a burst of $(+)$ energy, both of 
which are directed at a mirror. By ``rocking'' the mirror, we might try to deflect 
the $(+)$ and $(-)$ energy off at different angles, directing the $(+)$ energy off to some 
distant corner of the universe and the $(-)$ energy away into our laboratory \cite{PD}. 
However, this would mean that an 
observer very far from the mirror could intercept only the $(-)$ energy. 
This situation would violate the QIs, which are known to hold for 
{\it all} quantum states and all inertial observers in Minkowski 
spacetime. Therefore, what must happen is that the act of rocking the mirror 
creates compensating pulses of $(+)$ energy. 

\section{Application of the QI Bounds in Curved Spacetime}
\label{sec:CST} 
The QI bounds, although originally derived for flat spacetime, should 
hold in a curved spacetime and/or one with boundaries for 
``sufficiently small'' sampling times. The physical interpretation of 
this assumption is that we do not have to know about the large-scale 
curvature of the universe or the presence of distant boundaries in 
order to use flat spacetime QFT to accurately predict the outcome 
of a local (i.e., laboratory-scale) experiment.\footnote{Some possible 
caveats will be discussed later.}

\subsection{Common Misconceptions about QI Bounds}
At this point we take the opportunity to dispel some misconceptions about 
the QI bounds which appear from time to time. These are:

1) {\it QIs hold only in flat spacetime.} In fact, QI bounds have now been 
proven in a number of curved spacetimes as well \cite{PF971,PFGQI,FT,FLAN2,Fewster,FP}. 
In the short sampling time limit of these bounds, one does in fact recover 
the original flat spacetime bounds \cite{PF971,PFGQI}. 

2) {\it QIs break down at horizons \cite{Hay}.} This is not quite true. 
The QI bound depends on the observer you pick. One always gets a true bound, but 
not necessarily a strong bound. If you pick a ``bad'' observer (such as a static observer 
near the horizon of a black hole in the Unruh vacuum state), you get a very weak bound. 
However, if you choose an observer who falls freely through the horizon you can get a 
perfectly good and much stronger bound. (This is nicely shown in a recent 
paper by Flanagan \cite{FLAN2} for 2D black holes.) Both bounds are true, but the second 
provides a stronger constraint on the $(-)$ energy distribution.

3) {\it QIs don't hold for the Boulware vacuum state.} The issue of the Boulware vacuum state 
(as well as other instances where the energy density becomes singular) was discussed by Pfenning, 
Ford, and the present author \cite{FPR}. 
In this case, $r = 2M$ acts rather like a boundary at which the energy density 
becomes singular. Contrast this with the case of the Unruh or Hartle-Hawking vacuum states, 
where $r = 2M$ is a surface through which one can fall without encountering any infinite 
stress-energies. Hence, in the Boulware vacuum state, the relevant distance is not the proper 
radius of  curvature but the proper distance to the boundary. In fact, one can get a QI bound 
in this case using compactly supported sampling functions, provided that the tail of the sampling 
function does not intersect $r=2M$. 

4) {\it The tails of the sampling function can pick up large $(-)$ energy contributions from far away.} 
Although the original bounds were proven using a Lorentzian sampling function, QIs have since 
been proven using compactly-supported sampling functions \cite{FE,FLAN1,FLAN2,FP}. 
The validity of the bounds is not limited by 
sampling functions with tails. In fact, as mentioned previously, the currently known bounds 
hold for arbitrary (smooth) sampling functions.

5) {\it QIs don't hold for the Casimir effect \cite{OG}.} If the sampling time is 
chosen to be small compared to the proper distance between the 
Casimir plates, then one does get a QI bound \cite{FRWH}. Furthermore, 
the {\it difference} between the expectation value of the energy density in an 
arbitrary state and in the Casimir vacuum state {\it does} satisfy a QI 
bound \cite{FR95,P-Thesis}. The physical interpretation of this 
``difference inequality'' is that, although it is possible to depress 
the energy density below the vacuum Casimir value, it cannot be made 
arbitrarily negative for an arbitrarily long time. In the infinite 
sampling time limit one finds that {\it this difference of expectation 
values satisfies the AWEC, even though the Casimir vacuum 
energy by itself does not}.

6) {\it Flat space QIs applied to curved spacetime are incompatible with the existence 
of Hawking radiation \cite{Hay}.} In fact, the flat space QIs applied to an observer 
freely falling across the horizon of a Schwarzschild black hole in the Unruh vacuum state, 
which represents a black hole evaporating into empty space, do give a sensible bound 
\cite{FLAN2,FRunpub}.

\section{Wormholes and Time Machines} 
\label{sec:WHs}
Some time ago it was shown that two wormholes in motion relative to one 
another could be used, in classical GR, to build a time machine \cite{MT}. 
Shortly after, Morris, Thorne, and Yurtsever \cite{MTY} demonstrated that 
the same thing could be accomplished using just a single wormhole, by 
fixing the location of one wormhole mouth and moving the other.  

In 1992, Hawking gave a general proof to the effect that $(-)$ energy is 
{\it always} required to build a time machine in a ``finite region'' 
of spacetime (more precisely, for Cauchy horizons 
which are compactly generated) \cite{H92}. In that paper he also proposed his 
{\it chronology protection conjecture}: 
``the laws of physics will always prevent the occurrence of time machines.'' This 
conjecture is still unproven. For a good recent review of the current status of time machines, 
see the article by Visser \cite{V-HBV}. 

\subsection{Implications of QIs for Wormholes and Warp Drives} 
\label{sec:QI-WH-WD}
It was shown in 1996 by Ford and the present author \cite{FRWH} that the QI bounds imply: 
{\it either} a wormhole must have a throat size, $r_0$, no larger than about a few 
thousand Planck lengths ($l_p = 10^{-35} \rm m$), {\it or}, (typically) 
the $(-)$ energy must be confined to an extremely thin band around the throat. 
Similar problems afflict warp drive spacetimes \cite{A,PFWD,ER,K98}. 
(For some recent claims to the contrary, see Krasnikov \cite{K03}.) 
Yet another misconception about QIs is that they imply that 
wormholes can only be slightly larger than Planck scale \cite{FW,K00}. 
Ford and I have never claimed this. 
In our original 1996 paper, we were careful to make the ``either-or'' statement 
above. Our results are consistent with the semiclassical wormhole solutions of 
Hochberg, Popov, and Sushkov \cite{HPS} (their ``large throat'' wormholes have 
$r_0 \sim 200-300 l_p$), and also of Krasnikov \cite{K03,K00} (the latter has a 
large discrepancy in the length scales which characterize his wormhole). 

\subsubsection{Ways to Avoid This Conclusion?}
What are some ways one might think of to avoid this conclusion? \break
1) {\it N fields?} Superpose the effects of N fields, each of which satisfies the QI bound. 
The bound, Eq.~(\ref{eq:QI}), then becomes: 
\begin{equation}
\hat\rho \geq -\,\frac{N \,C \, \hbar}{c^3 \, \tau_0^4}\,.
\end{equation}
In practice, N must be {\it enormous} to have any significant effect \cite{FRWH}. 
For example, to get $r_0 = 1 {\rm m}$ for a wormhole, one needs about 
$10^{62}$ fundamental fields (or a few fields for which $C \gg 1$, 
i.e., $C$ of order $10^{62}$). 

2) {\it Interacting fields?} To date, the QIs have only been proven for free 
quantum fields. The interacting case is technically much more difficult. However, 
one might expect that for sampling times which are small compared to the 
interaction timescale, the usual free field QIs should apply. However, Olum and Graham \cite{OG} 
have recently constructed a 2+1 dimensional model of two interacting scalar fields, one of which 
is similar to a classical domain wall, where the energy density can be {\it static and 
negative} in certain regions. Whether it can be made static and negative over an 
{\it arbitrarily large} spatial region is another matter. It seems more likely that, 
as in the case of the Casimir effect, the larger the magnitude of the $(-)$ energy density, 
the ``narrower'' the region of space it will be allowed to occupy.

3) {\it $(-)$ energy cosmic strings?} Cosmic strings concentrate large energy densities 
in very small regions of space. If $(-)$ energy density cosmic strings can exist, then they 
might be used as sources for macroscopic wormholes \cite{V-book} with the 
requisite large discrepancies in length scales implied by the QIs. Unfortunately, 
the models of cosmic strings derived to date all have $(+)$ energy densities \cite{VS}.

4) {\it Classical fields with $(-)$ energy density?} Recently Barcelo and 
Visser have constructed wormhole solutions using classical non-minimally coupled 
scalar fields as the negative energy source \cite{BV1,BV}. This possibility 
will be discussed in a later section.   

5) {\it Dark energy with a $(-)$ energy density?} 
Recently there have been several models and arguments which propose that the dark energy 
driving the acceleration of the universe may have an equation of state which allows $(-)$ 
energy densities, at least as seen by some observers \cite{PKV,MMOT,CHT}. 
This $(-)$ energy component would grow with time and would not seem 
to be limited by the QIs. If true, this would seem rather puzzling and a bit disturbing. 
This would seem to allow the possibility that an observer holding a box could see 
it fill up with an arbitrarily large amount of negative energy as the universe expanded. 

6) {\it Ford-Svaiter parabolic mirror example?} Ford and Svaiter 
have recently shown that vacuum fluctuations can be focused by a cylindrical 
parabolic mirror \cite{FS1,FS2}. 
The sign of the vacuum energy density can be made negative at the focal line of the mirror. 
In their model it appears that the smallest relevant length scale is not the proper distance 
to the mirror (which can be arbitrarily far away), but instead the distance to the focal point 
of the mirror, which is much smaller. If this result is indeed confirmed by further 
calculations, then the simple prescription that the flat space QIs should hold for 
sampling times small compared to the smallest proper radius of spacetime curvature or 
the smallest proper distance to a boundary would need to be refined.

\section{QIs: Some Recent Developments} 
\label{sec:QI-RD}
\subsection{``No-go'' results} 
\label{sec:NoGo}
Given the success of finding QIs averaged along timelike worldlines, 
a natural question arises as to whether there exist similar QIs which involve 
averaging along spatial directions alone. Furthermore, are there ``spacetime-averaged'' 
QIs, which reduce to the usual worldline QIs in the case where all the spatial sampling 
lengths go to zero, and which reduce to spatially averaged QIs when the 
sampling time goes to zero? Spacetime-averaged QIs do exist in 2D spacetime, 
and reduce to purely spatial QIs in the zero sampling time limit \cite{FLAN1,TR-unpub}.

By contrast, it has been proven by means of a counterexample that 
no purely spatial QIs exist in 4D, even in Minkowski spacetime \cite{FHR,F-MGM}. 
However, recent evidence implies 
that spacetime QIs probably do exist in 4D (i.e., stronger than what one 
would get by simply combining the worldline QIs of 
many observers in a region) \cite{Funpub,DF}. 

Here we briefly trace the construction of the counterexample mentioned above. 
We work with the massless minimally coupled scalar field 
and choose our set of quantum states to be a superposition of vacuum plus 
two-particle states, of the form 
\begin{equation} 
|\psi \rangle = N \left[ |0\rangle + \sum_{{\bf k}_1,{\bf k}_2}  
c_{{\bf k}_1,{\bf k}_2}\, |{{\bf k}_1,{\bf k}_2}\rangle \right] \,, 
\label{eq:state} 
\end{equation} 
where we assume that the $c_{{\bf k}_1,{\bf k}_2}$ are symmetric.  
Here $|{{\bf k}_1,{\bf k}_2}\rangle$ is a two-particle state containing 
a pair of particles with momenta ${\bf k}_1$ and ${\bf k}_2$, and the 
normalization factor is 
\begin{equation} 
N = \left[1 + \xi \sum_{{\bf k}_1,{\bf k}_2} |c_{{\bf k}_1,{\bf k}_2}|^2  
\right]^{-\frac{1}{2}} \,, 
\label{eq:norm_a} 
\end{equation}
where $\xi = 2$ if ${\bf k}_1 \not= {\bf k}_2$ and 
$\xi = 1$ if ${\bf k}_1 = {\bf k}_2$. 
The $c_{{\bf k}_i,{\bf k}_j}$ are chosen such that 
pairs of particles in the two-particle component of the state 
have {\it nearly but not exactly opposite momenta}, in the high frequency limit. 
Although their momenta nearly cancel, their energies do not. This has 
the effect that spatial oscillations of the stress-tensor are severely damped, 
while temporal oscillations are greatly enhanced. In 
Minkowski spacetime, at a {\it single instant in time} 
the energy density in an arbitrarily large compact region of space can be made 
approximately constant and arbitrarily negative. However, the worldline QIs are 
still enforced even in this state, due to the large time oscillations in the stress tensor. 
Note that our result might appear similar to that of the Olum and Graham interacting scalar field 
model discussed earlier. It is thus important to emphasize here that, in our example, the energy 
density can be made arbitrarily negative over an arbitrarily large compact spatial region, 
{\it only at a single instant in time.} As conjectured earlier, it seems more likely that 
in the Olum-Graham example, large magnitude static $(-)$ energy densities 
will be constrained to narrow regions of space, as in the Casimir effect.

A similar construction shows that no QIs exist along null geodesics in 4D flat spacetime, 
although they do exist in 2D \cite{FewR}. More specifically, it 
can be shown that for certain states, the quantity $\langle T_{\mu\nu} K^{\mu} K^{\nu} \rangle$ 
sampled along a null geodesic can be made arbitrarily negative over a finite length of affine 
parameter, where $K^{\mu}$ is the affinely parameterized tangent vector to the null 
geodesic. (However, the ANEC still holds even in these states.) This result dashes hopes of 
proving singularity theorems which simply replace the ANEC with QIs along null geodesics. 
However, the quantity $\langle T_{\mu\nu} K^{\mu} K^{\nu} \rangle$ 
sampled along a {\it timelike} geodesic {\it does} obey a QI bound. 
Whether this bound is useful for proving global results in GR remains 
to be seen.

\section{Wormholes: Some Recent Developments}
\label{sec:WH-RD}
Visser, Kar, and Dadhich \cite{VKD} have recently argued that, 
in principle, it is possible to construct wormholes with arbitrarily 
small amounts of exotic matter. These wormholes are not without 
problems however. Most importantly, they can be severely constrained 
by the QIs, using the same arguments given earlier \cite{FRWH,TR2-unpub}. 
There are also some practical difficulties. 
The smaller the amount of exotic matter used in these wormholes, 
the closer they are to being vacuum Schwarzschild wormholes. Hence: 
(1) the smaller the amount of exotic matter, the longer it will take an observer 
to traverse the wormhole as measured by clocks in the external universe. 
(Perhaps one could counter this by moving the wormhole mouths around?)
(2) The smaller the amount of exotic matter, the more prone the wormhole is 
to destabilization by even very small amounts of infalling positive matter, 
since this matter will be enormously blueshifted by the time it reaches the throat. 
Most recently, Kuhfittig has constructed wormhole solutions which are 
claimed to satisfy both the traversability constraints and the QI bounds, 
and which require only arbitrarily small amounts of exotic matter \cite{Kuh}. 
Hong and Kim have discussed the possibility of associating a negative temperature 
to wormholes as a result of the presence of the exotic matter \cite{HK}.

Barcelo and Visser \cite{BV1,BV} have proposed classical non-minimally coupled scalar fields 
as sources of exotic matter for wormhole maintenance. 
Since such classical fields would not be subject to the QIs, one might expect 
to be able to use them to produce other macroscopic effects, such as violation 
of the generalized second law (GSL) of thermodynamics. However, an attempt to manipulate 
classical non-minimally coupled scalar fields to shoot a $(-)$ energy flux 
into a black hole with the goal of violating the GSL fails. The proof that the GSL 
is preserved \cite{FR01} depended crucially on the assumption that 
$(1-8 \pi \xi \, \phi^2) > 0$ everywhere, where $\xi$ is the coupling 
parameter and $\phi$ is the scalar field. This condition, which is needed to 
even define the transformation from the ``Jordan frame'' 
to the ``Einstein frame'', is violated in the Barcelo-Visser 
wormhole solutions.\footnote{The theory of GR with a non-minimally coupled scalar 
field is mathematically equivalent to the theory of GR plus a minimally 
coupled scalar field. The use of the word frame here is perhaps misleading; the   
transformation between the two is {\it not a coordinate   
transformation}, but rather a field redefinition which mixes   
the nature of the scalar and gravitational fields.} 
Although it has not been proven explicitly, one strongly suspects 
that if the condition is violated, then one can construct examples which 
violate the GSL. 

Nonetheless it is quite interesting that, even for the case of 
relatively weak fields and coupling constants: (a) one can get large 
sustainable, albeit temporary, classical $(-)$ energy fluxes (even in flat spacetime!), 
and, (b) such fluxes do not violate the GSL. 

At this point, the author must admit to a philosophical prejudice. 
If classical fields which violate the WEC really do exist, 
then why is this true of only certain classical fields? The conformally 
coupled scalar field is often deemed ``physically reasonable'' in part 
because it faithfully mimics certain behaviors of the EM field. So why 
then does the classical EM field obey the WEC, but the classical C-CS 
field does not? Also, why does Nature appear to try so hard to limit 
quantum violations of the energy conditions (in the form of QIs) if 
they can easily be violated classically? 

\section{Conclusions} 
\label{sec:C}
Much progress has been made during the last decade in deriving ``quantum inequality (QI)'' 
constraints on the magnitude and duration of $(-)$ energy densities in quantum 
field theory. QIs have now been proven for a number of 
(free) quantum fields including the EM field, the massive and massless minimally 
coupled scalar fields, and the Dirac field in both 2D and 4D spacetimes. QIs have been 
proven using arbitrary (smooth) sampling functions, including compactly-supported sampling 
functions. They have also been proven in a number of curved spacetimes. QIs place some severe 
constraints on wormhole and warp drive spacetime geometries. On the other hand, two notable ``no-go'' 
results have recently been proven. No purely spatial QIs exist in 4D 
spacetime (although they do exist in 2D). Similarly, no QIs exist along null 
geodesics in 4D (although they do exist in 2D). There is evidence that non-trivial 
spacetime-averaged QIs exist in 4D, although 
no simple analytic forms have yet been written down.

\subsection{Future Directions}
\label{sec:FD}
The QIs constrain the magnitude and duration of $(-)$ energy densities relative to 
those of an underlying reference vacuum state. A deeper understanding of the possible existence 
of constraints on vacuum $(-)$ energy densities is needed. Although the QIs do provide {\it some} 
bounds on vacuum $(-)$ energy densities \cite{FRWH}, more 
work is required. Typically, large static negative energy densities associated with vacuum states 
are concentrated in narrow spatial regions, e. g., between a pair of Casimir plates with small 
separation or the region near $r=2M$ in the Boulware vacuum 
\footnote{The energy density seen by static observers is everywhere negative in the Boulware vacuum state. 
However, its magnitude is very large only extremely close to $r=2M$.}. Must this always be true? 
Can such $(-)$ vacuum energy be ``mined'' or otherwise extracted to produce gross macroscopic effects \cite{WI}? 
To give a somewhat crude example, could one somehow isolate and collect some of the $(-)$ energy 
in the Boulware vacuum, take it back to our laboratory, and use it to construct a traversable 
wormhole? One tends to suspect that the answer is no, but a rigorous proof to that effect has not yet 
been presented. This may be related to the idea of ``passive states'' and the 
recently discovered connection between thermodynamics, QIs, and microlocal analysis \cite{Kuck,FewV}.

The QIs have been proven to date only for free fields. It would be 
rather surprising if the inclusion of interactions drastically modified 
the current picture. Otherwise, one might worry that this might lead to 
instabilities in the free field theory as well. 
However, given the Olum-Graham results, further study of interacting fields is needed. 

Further analysis and understanding of the Ford-Svaiter parabolic mirror vacuum 
fluctuation focusing model is warranted. It is important to confirm whether 
there might generally be additional relevant geometrical length scales to consider 
when applying the flat space QIs to curved space or spaces with boundaries, other 
than just the smallest proper radius of curvature or the smallest proper distance 
to a boundary. Similarly, the accelerating universe models which involve $(-)$ 
energy densities should be scrutinized as possible examples of a field 
which does not obey the QIs.

Although there is evidence for the existence of spacetime-averaged QIs in 4D, no explicit 
simple analytic forms have yet been written down. This is a key unsolved problem. 
The techniques used to prove the worldline QIs do not seem to easily generalize to 
the spacetime-averaged case. Even simple explicitly constructible states involving 
$(-)$ energy densities, e.g., squeezed vacuum states of a massive scalar field, 
can exhibit a subtle intertwining of the $(-)$ with the $(+)$ energy. 
How generic is this behavior? One can constrain large classes of possible 
spatial distributions of $(-)$ energy in flat spacetime, using 
simply the AWEC and the QIs \cite{BFR}. This approach may be complementary to 
spacetime-averaged QIs. How far can this program be pushed?

The QIs are bounds on averages of the expectation value of the stress tensor. Such an analysis 
does not take into account $(-)$ energy densities associated with stress tensor fluctuations, 
which could be important for areas such as inflation \cite{BorV,SW,Vach}. 

More work needs to be done in understanding the $(-)$ energy densities associated 
with classical non-minimally coupled scalar fields. How seriously should one 
take classical fields with $(-)$ energy? If such fields are truly physical, then why 
does Nature bother to enforce QIs at all?

The fascinating mysteries and subtleties of negative energy should keep us 
all busy for a while yet.

\section*{Acknowledgments}
The author would like to thank Larry Ford for useful comments. 
This work was supported in part by NSF Grant No. PHY-0139969.

\end{document}